
To: <HEP-TH@XXX.lanl.gov>
Passed-Date:  MON, 12 JUL 1993 21:07 EXP
Passed-From:  <LHJ5551@KRYSUCC1>
Passed-To:    <HEP-TH@XXX.LANL.GOV>

\input harvmac
%
%
\def\today{\ifcase\month\or
   January\or February\or March\or April\or May\or June\or
   July\or August\or September\or October\or November\or December\fi
   \space\number\day, \number\year}
%
%
%
%
%
\def\dooley{\centerline{Hyuk-jae Lee\footnote{$^*$}
{Bitnet: lhj5551@krysucc1}}\bigskip
\centerline{{\it Department of Physics, Yonsei University}}
\centerline{{\it Seoul 120-749, Korea}}\bigskip}
\def\submit{\baselineskip=20pt plus 2pt minus 2pt} 
\def\a{\alpha}           \def\c{\chi}       \def\d{\delta}
     \def\f{\phi}       \def\F{\Phi}
\def\vf{\varphi}  \def\g{\gamma}           
\def\l{\lambda}   \def\L{\Lambda}             
\def\r{\rho}          \def\o{\omega}     \def\O{\Omega}
\def\p{\psi}      \def\P{\Psi}        \def\s{\sigma}     
     \def\t{\tau}       
\def\x{\xi}               \def\w{\wedge}
\def\e{\eta}      \def\Th{\Theta}  \def\td{\tilde}
%

\def\CA{{\cal A}}   
 \def\CF{{\cal F}}  \def\CG{{\cal G}}
\def\CH{{\cal H}}   
\def\CO{{\cal O}}   
   
\def\CW{{\cal W}}   \def\CS{{\cal S}}
\def\CE{{\cal E}} \def\CO{{\cal O}}
%
\def\rd{\partial}

\def\darr#1{\raise1.5ex\hbox{$\leftrightarrow$}\mkern-16.5mu #1}
\def\Ha{{1\over2}}

\def\Fr#1#2{{#1\over#2}}

%
\def\cmp#1#2#3{, Comm.\ Math.\ Phys.\ {{\bf #1}} {(#2)} {#3}}
\def\pl#1#2#3{, Phys.\ Lett.\ {{\bf #1}} {(#2)} {#3}}

\lref\singer{L. Baulieu and I. M. Singer, Nucl. Phys. (Proc. Suppl)
            {\bf 5B}(1988)12}
\lref\ken{H. Kanno, Z. Phys.{\bf C43}(1989)43}
\lref\ati{M. F. Atiyah and I. M. Singer, Proc. Natl. Acad. Sci. USA
{\bf 81}(1984)2597}
\lref\wit{E. Witten\cmp{117}{1988}{353}}
\lref\lab{J. M. F. Labastida and M. Pernici\pl{B212}{1988}{56}}
\lref\bra{S. B. Bradlow\cmp{135}{1990}{1}}
\lref\kob{S. Kobayashi,{\it Differential Geometry of Complex Vector Bundles},
Princeton University Press (1987)}
\lref\sch{F. A. Schaposnik and G. Thompson\pl{224}{1989}{379}}
\lref\cha{G. Chapline and B. Grossman\pl{223}{1989}{336}}
\lref\bau{L. Baulieu and B. Grossman\pl{214}{1988}{223}}
\lref\lee{H. J. Lee,Preprint YUMS-93-16, SNUTP-93-45}
\lref\brai{S. B. Bradlow, J. Differential Geometry, {\bf 33}(1991)169}
\lref\park{J. S. Park, {\it $N=2$ Topological Yang-Mills Theory on
Compact K\"ahler Surfaces},Preprint}

\Title{\vbox{\baselineskip12pt\hbox{YUMS 93-17}\hbox{SNUTP 93-49}}}
{\vbox{\centerline{Topological Field Theory of Vortices}
    \vskip2pt\centerline{over Closed K\"ahler Manifolds}}}
\dooley
\vskip 0.3 in
\submit
By dimensional reduction, Einstein-Hermitian equations of $n+1$ dimensional
closed K\"ahler manifolds lead to vortex equations of $n$ dimensional
closed K\"ahler manifolds. A Yang-Mills-Higgs functional to unitary bundles
over closed K\"ahler manifolds has topological invariance by adding the
additional terms which have ghost fields. Henceforth we achieve the matter
(Higgs field) coupled topological field theories in higher dimension.

\Date{}

\newsec{Introduction}
At Yang-Mills gauge theory on four manifolds Witten\wit\ set up the
relativistic field theory with a global fermionic symmetry similar to
BRST symmetry. This field theory has been called by topological quantum
field theory(TQFT). TQFT could be constructed by BRST gauge fixing on
the Yang-Mills instantons space of anti-self dual equations\singer\lab.
That anti-self dual equations arise as minimizing conditions for gauge
invariant functional. In two and tree manifolds as minimizing conditions
we can choose the self dual and anti-self dual Yang-Mills equations, and vortex
equations and monopole equations which have Higgs fields.
Some authors showed that by choosing the solutions space of those
equations as gauge fixing condition TQFT can be constructed\sch\cha\bau.

If one considers a closed K\"ahler manifold and considers unitary
connection on a complex bundle, the equations for the minima are the
Einstein-Hermitian equations. The conditions for the existence of
solutions of the Hermition-Eistein equations can be related to the
stability of the holomorphic bundle. On the complex surface one can set
up TQFT\lee\park.

Bradlow\bra\brai\ showed that existing the global minima of the
Yang-Mills-Higgs
is extended into arbitrary dimensional closed K\"ahler manifolds.
and studied the necessary and sufficient conditions for the existance of
solutions to the vortex equations.

We will show that the vortex equations is obtained by demensional
reduction of Einstein-Hermitian equations. In this scheme we will construct
topological field theory over vortices, that is, a Yang-Mills-Higgs functional
to unitary bundle over closed K\"ahler manifolds has fermionic symmetry.

In section 2, we review the moduli space on closed K\"ahler manifolds and
intend to obtain the vortex equations from dimensional reduction  $n+1$
into  $n$ dimensional manifolds. In section 3, we show that
the Yang-Mills-Higgs
theory has fermionic symmetry (topological symmetry) over vortices
space. In section 4, on 2-dimension K\"ahler manifolds we construct the
topological invariant observables.

\newsec{The moduli space of Yang-Mills Theory on closed K\"ahler Manifolds}
\subsec{The Einstein-Hermition Structure}
Let X be a closed K\"ahler manifold of complex dimension $n$ and let $\o$
be the K\"ahler form. Let $E$ be a complex vector bundle of rank $n$ over
$X$. We will consider the vector bundle $E$ to be endowed with a fixed
hermitian metric $h$. The Hermitian metric $h$ in $E$ is a smooth field of
Hermitian inner products in the fibers of $E$ such that
\eqn\fib{\eqalign{
& h(\x,\e)\ \ is\ linear\ in\ \x,\quad where\ \x,\e\in E_x,\cr
& h(\x,\e)=\overline{h(\e,\x)},\cr
& h(\x,\x)>0 \quad \x\neq 0,\cr
& h(\x,\e)\ is\ a\ smooth\ function\ if\ \x\ and\ \e\ are\ smooth\
sections.\cr}
}
We call $(E,h)$ an Hermitian vector bundle.
Let $\CA(h)$ denote connections on $E$ that are unitary with respect to $H$.
We use the following;\hfil\break
$\O^{p,q}(X)=$ the space of $(p,q)$-forms over $X$,\hfil\break
$\O^{p,q}(X,E)=$ the space of $(p,q)$-forms with values in $E$,\hfil\break
$\O^{p,q}(X,End\ E)=$ the space of $(p,q)$-forms with values in
the endomorphism bundle of $E$.\hfil\break
All the spaces carry hermitioan metric induced by K\"ahler metric
on $X$ and the metric on $E$. So we lead to identifications $E\approx E^*$
and $E\otimes E^* \approx End\ E$.
We define the dual operator
\eqn\lli{
\bar\ast:\O^{p,q}\to \O^{n-p,n-q}.
}
We define an inner product in the space of $(p,q)$-forms on $X$ by setting
\eqn\lei{
(\f,\p)=\int \f\w\bar\ast\p.
}
On K\"ahler manifold with K\"ahler form $\o$, we can define a map
\eqn\leii{
L:\O^{p,q}(X,E)\to \O^{p+1,q+1}(X,E)
}
by
\eqn\leiii{
L(\a)=\a\w\o.
}
The adjoint of this map for the inner product \lei\ is given by $\L$
such that
\eqn\leiv{
\L:\O^{p,q}(X,E)\to \O^{p-1,q-1}(X,E).
}

Let $d_A=d+A$ be a connection in $E$. We can split of $d_A$ into
$\rd_A+\bar\rd_A$ on complex vector bundle over $X$ where
\eqn\yyi{\eqalign{
&\rd_A:\O^{p,q}(X,E)\to\O^{p+1,q}(X,E),\cr
&\rd_A\equiv \rd+A'\cr}
}
and
\eqn\yyii{\eqalign{
&\bar\rd_A:\O^{p,q}(X,E)\to\O^{p,q+1}(X,E),\cr
&\bar\rd_A\equiv \rd+A''.\cr
}}

We will denote $F$ the curvature of $d_A$,
i.e. $F=d_A\circ d_A\in\O^2(X,End\ E)$.
Then we can decompose as
\eqn\yyiii{
F=F^{2,0}+F^{1,1}+F^{0,2},
}
where
\eqn\yyv{\eqalign{
F^{2,0}\equiv \rd_A\circ\rd_A\in\O^{2,.0}(X,End\ E),\quad
F^{0,2}\equiv \bar\rd_A\circ\bar\rd_A\in\O^{0,2}(X,End\ E),\cr
F^{1,1}\equiv (\rd_A\circ\bar\rd_A+\bar\rd_A\circ\rd_A)\in\O^{1,1}(X,End\ E).
}}
$F^{1,1}$ can be decompose into
\eqn\uuu{
F^{1,1}=F^{1,1}_0+F^0\cdot\o,
}
where $F^{1,1}_0$ consists of forms orthogonal to $\o$.
A connection $d_A$ is called integrable if $(\bar\rd_A)^2=0$. If $d_A$
is a connection in $E$ such that $(\bar\rd_A)^2=0$, $\bar\rd_A$ defines
a holomorphic structure on $E$. A section $\x\in\O^0(X,E)$ is called
holomorphic if and only if $\bar\rd_A\x=0$. The complex vector bundles
which admit holomorphic structures is called vector bundle over a
complex manifold $M$. From now on we shall study holomorphic vector
bundle.

Let $\CH (h)$ denote the subset of $\CA(h)$ consisting of
$d_A=\rd_A+\bar\rd_A$ such that $\bar\rd_A \circ \bar\rd_A=0$ and let
$\CE(h)$ denote the subset of $\CH(h)$ consisting
of Einstein-Hermitian connections $d_A$, i.e.
\eqn\ein{
\CE(h)=\{d_A\in\CH(h)|i\L F=cI_E\}
}
where $c$ is a constant and $I_E\in\O^0(x,End\ E)$ is the identity
section.

\subsec{Vortices}
Let $\O^0 (X,E)$ denote the smooth sections of $E$. We consider integrable
unitary connections belonging to $\CA^{1,1}(h)$ such that
\eqn\tty{
\CA^{1,1}(h)=\{d_A \in \CA^{1,1}|F^{0,2}=0\}
}
and the pairs $(d_A,\F)$ in $\vf\subset\CA^{1,1}\times\O^0(X,E)$
where
\eqn\tti{
\Th=\{(\d_A,\F)\in \CA^{1,1}\times\O^0(X,E)|\bar\rd_A\F=0\}.
}
The complex gauge group acts on both these spaces by
\eqn\ttii{\eqalign{
& g(\bar\rd_A)=g\circ\bar\rd_A\circ g^{-1},\cr
& g(\F)=g\F.\cr}
}
This vector valued function $\F$ is called a Higgs field.

Among the triple $(\bar\rd_A,\F,h)\in\Th$, if it satisfys the equations,
\eqn\ttii{\eqalign{
& \bar\rd_A\F=0,\cr
& i\L F_{\bar\rd_A,h}+\Ha\F\otimes\F^{*h}-\Ha\t I_E=0,\cr}
}
we will call the metics $h$ $\t$-Hermitian-Yang-Mills-Higgs metrics.
Here $F_{\bar\rd_A,h}$ is the curvature of the metric connection compatible
with $\bar\rd_A$ and $h$ and $\F^{*h}$ denote that the adjoint is taken
with respect to the metric $h$.

Bradlow\bra\ presented that the triple $(\bar\rd_A,\F,h)$ satisfied
Eq.\ttii\ are global minima of the Yang-Mills-Higgs action.

\subsec{The Dimensional Reduction}
We consider the $n+1$ dimensional closed K\"ahler complex manifold.
We will show that with the dimensional reduction $n+1$ dimension to $n$
dimension manifold, the Einstein-Hermitian condition on the $n+1$ dimensional
manifold leads to the Hermitian-Higgs condition on the $n$ dimensional
manifold.

The connection is described by
\eqn\tyi{
\td A=A'_1 dz_1+\cdots+A'_{n+1}dz_{n+1}+A''_1 d \bar z_1+\cdots
+A''_{n+1}d\bar z_{n+1}.
}

We now assume that the connections $A_i$ are independent of the coordinates,
$z_{n+1}$ and $\bar z_{n+1}$, and define functions of $(z_1,\cdots,z_n,
\bar z_1,\cdots, \bar z_n)$. Thus over $n$ dimensional manifold we can
define the connection
\eqn\tyii{
A=A'_1 dz_1+\cdots+A'_n dz_n +A''_1 d \bar z_1+\cdots+A''_n d \bar z_n,
}
and $A'_{n+1}$ and $A''_{n+1}$ relabel as $\F^*$ and $\F$ which are independent
of $z_{n+1}$ and $\bar z_{n+1}$.
We denote $\td F$ the curvature form of vector bundle over $n+1$ dimensional
manifold and $F$ the curvature form of vector bundle over $n$ dimensional
manifold.

We assume that the curvature $\td F$ of the vector bundle over the $(n+1)$
dimensional base manifold $M$ can be rewritten by
\eqn\yyvi{
\td F=F^{2,0}+F^{0,2}+F^{1,1}+\rd_A\F^*\w dz_{n+1}+\bar\rd_A\F\w d\bar z_{n+1}+
\F^* dz_{n+1}\w\F d\bar z_{n+1}}
where $F$'s are the curvature of vector bundle over the $n$ dimensional
base manifold $X$ and $\F$ independent of $\{ z_{n+1},\bar z_{n+1}\}$
relabels the $(n+1)$th component of the connection.

The integrable condition $\td F^{0,2}=0$ on the $n+1$ dimensional manifold
leads to
\eqn\tyiii{\eqalign{
& F^{0,2}=0,\cr
& \bar\rd_A \F=0.\cr}
}
on the $n$ dimensional manifold.
The first equation denotes the integrable condition on the $n$ dimensional
manifold and the second equation describes that $\F$ has holomorpic condition.
For the Einstein-Hermitian condition on the $n+1$ dimensional manifold,
$i\L \td F=cI_E$ leads to vortex equation,
\eqn\tyiv{
i\L F+\Ha \F \otimes \F^*= \Ha\tau I_E
}
on the $n$ dimensional manifold
where we substitute $\tau \over 2$ for $c$.
In fact,
\eqn\hhh{
\L(\F^* dz_{n+1}\w\F d\bar z_{n+1})=\Ha |\F|^2_h
}
where $\L$ is adjoint operator of $L$ which is defined locally by
\eqn\ppp{
L\f=\f\w(\Fr{i}{2}dz_{n+1}\w d\bar z_{n+1})
}

Then we can find that the connections space $\CE(h)$ on the $n+1$ dimensional
manifold leads to the stable pairs space,
$\CA^{1,1}(h)\times\O^0(X,E)$.

\newsec{Topological Quantum Field Theory}
We can obtain the following transformation laws from the curvature formula
and Bianchi identity for the universal bundle over $M\times\CA/\CG$ for
$n+1$ dimensional manifold $M$\ken\ati\singer;
\eqn\li{\eqalign{
s\td A' & =\td\p,\cr
s\td A'' & =\td{\bar\p},\cr
s\td\p & =-\rd_A \f,\cr
s\td{\bar\p} & =-\bar\rd_A\f,\cr
s\f &=0,\cr}  `
}

Like Eq.\tyi, we assume that the ghost fields $\td\p$ and $\td{\bar\p}$ are
independent of the coordinates $z_{n+1}$ and $\bar z_{n+1}$ and define
functions of $(z_1,\cdots,z_n,\bar z_1,\cdots,\bar z_n)$. The $(n+1)$-th
components $\td\p_{n+1}$ and $\td{\bar\p}_{n+1}$ relabel as $\a$ and $\bar\a$
which are independent of $z_{n+1}$ and $\bar z_{n+1}$.
Then we can arrive at the following transformation laws;
\eqn\tli{\eqalign{
s A' =\p,  \qquad  & s A'' =\bar\p,\cr
s\p =-\rd_A \f,\qquad & s\bar\p=-\bar\rd_A\f,\cr
s\F^* =\a, \qquad & s\F=\bar\a, \cr
s\a=-[\F^*,\f], \qquad & s\bar\a=-[\F,\f],\cr
s\f =0.\cr}
}

We need additional multiplets to write Lagrangian. So we introduce two
pairs $(\l,\eta)$, $(\x,\r)$ and $(\bar\c,\bar H)$ such that
\eqn\abc{\eqalign{
& s\l=\eta,\qquad\quad s\eta=[\f,\l],\cr
& s\x=\r, \qquad\quad s\r=[\f,\x],\cr
& s\bar\c=\bar H, \qquad\quad s\bar H=[\f,\bar\c].\cr}
}
The multiplets $(\bar\c,\bar H)$ are self-dual 2-forms and have the ghost
number $(-1,0)$. So the self-dual 2-forms $\c$ and $H$ can be written as
\eqn\abd{\eqalign{
& \bar\c=\c^{2,0}+\c^{0,2}+\c\o,\cr
& \bar H=H^{2,0}+H^{0,2}+H\o,\cr}
}
where $\c=\L\bar\c$ and $H=\L\bar H$.
The multiplets $(\x,\r)$ and $(\l,\eta)$ are zero forms and have the ghost
number $(-1,0)$ and $(-2,1)$, respectively.

In order to obtain the topological invariant action we are going to follow the
procedure of standard BRST gauge fixing. We concern the stable pair sector with
gauge coupled Higgs fields. So we choose the following gauge conditions
\eqn\ddd{\eqalign{
&F^{0,2}=0,\cr
&\bar\rd_A\F=0,\cr
&i\L F+\Ha \F\otimes\F^*=\Fr{\t}{2}I_E,\cr}
}
and the slice condition,
\eqn\bbb{
\L(\bar\rd_A \p -\rd_A \bar\p)+\Ha\F\otimes\a-\Ha\F^*\otimes\bar\a=0
}
which are orthogonalto the variations in the pair $(A,\F)$ that can be obtained
by a gauge transformations.

We can obtain the action as
\eqn\liii{\eqalign{
\CS=& s\int_M Tr[\c^{0,2}\w\bar\ast (F^{0,2}+\Fr{1}{16}H^{0,2})
+\x\w\bar\ast(\bar\rd_A\F+\Fr{1}{8}\r)\cr
& +\c(i\L F+\Ha\F\otimes\F^*-\Ha\t I_E+\Fr{1}{4}H)\o\w\bar\ast\o \cr
& +\l(\L(\bar\rd_A \p -\rd_A \bar\p)+\Ha\F\otimes\a-\Ha\F^*\otimes\bar\a)
\o\bar\ast\o]. \cr}
}
Then  by simlpe calculation one can find that
\eqn\lv{\eqalign{
\CS=&\int_M Tr[H^{0,2}\w\bar\ast (F^{0,2}+\Fr{1}{16} H^{0,2})
+\r\w\bar\ast(\bar\rd_A\F+\Fr{1}{8}\r)\cr
&+H(i\L F+\Ha\F\otimes\F^*-\Ha\t I_E+\Fr{1}{4}H)\o\w\bar\ast\o \cr
&+\c^{0,2}\w(\bar\ast\bar\rd_A\bar\p+\Fr{1}{16}[\f,\bar\ast\c^{0,2}])
+\x\w(\bar\ast\rd_A\p+\Fr{1}{8}[\f,\bar\ast\c^{2,0}]) \cr
& +\c\{ \L(\rd_A\bar\p +\bar\rd_A \p)+\Ha\bar\a\otimes\F^*+\Ha\F\otimes\a
+\Fr{1}{4}[\f,H]\} \o\w\bar\ast\o\cr
&+\eta\{\L(\bar\rd_A\p-\rd_A\bar\p)+\Ha\F\otimes\a-\Ha\F^*\otimes\bar\a\}
\o\bar\ast\o] \cr
& +\{\l((\bar\rd_A^*\rd_A +\rd_A^*\bar\rd_A)\f -\Ha\F\otimes[\F^*,\f]
+\Ha\F^*\otimes[\F,\f])\cr
& +\L(\bar\p\w[\l,\p]-\p\w[\l,\bar\p])\}\o\w\bar\ast\o].\cr}
}
$H$ is an auxiliary field important in closing the algebra but not
propagating. By using the Euer-Lagrange equation we can eliminating $H$,
one finds an action,
\eqn\llv{\eqalign{
\CS=&4||F^{0,2}||^2 +2||\bar\rd_A\F||^2 +||i\L F+\Ha\F\otimes\F^*
-\Ha\t I||^2\cr
&+\int_M Tr[\c^{0,2}\w(\bar\ast\bar\rd_A\bar\p+\Fr{1}{16}[\f,\bar\ast\c^{0,2}])
+\x\w(\bar\ast\rd_A\p+\Fr{1}{8}[\f,\bar\ast\c^{2,0}]) \cr
& +\c\{ \L(\rd_A\bar\p +\bar\rd_A \p)+\Ha\bar\a\otimes\F^*+\Ha\F\otimes\a
+\Fr{1}{4}[\f,H]\} \o\w\bar\ast\o\cr
&+\eta\{\L(\bar\rd_A\p-\rd_A\bar\p)+\Ha\F\otimes\a-\Ha\F^*\otimes\bar\a\}
\o\bar\ast\o] \cr
& +\{\l((\bar\rd_A^*\rd_A +\rd_A^*\bar\rd_A)\f -\Ha\F\otimes[\F^*,\f]
+\Ha\F^*\otimes[\F,\f])\cr
& +\L(\bar\p\w[\l,\p]-\p\w[\l,\bar\p])\}\o\w\bar\ast\o]].\cr}
}

The first three terms of the action \llv ,
\eqn\rrr{
4||F^{0,2}||^2 +2||\bar\rd_A\F||^2 +||i\L F+\Ha\F\otimes\F^* -\Ha\t I||^2
}
can be rewritten as
\eqn\rrri{
||F||^2 +||D\F||^2 +{1\over 4}||\F\otimes\F^*
-\t I||^2-\t\int_X iTrF \w\o^{n-1}-\int_X Tr(F\w F)\w\o^{n-2}.
}

Here this follows from the identities,
\eqn\rrrii{\eqalign{
|F|^2\o^n &=|\L F|^2\o^n+Tr(F\w F)\w\o^{n-2}+2(|F^{0,2}|^2 +|F^{2,0}|^2)
\o^n,\cr
<i\L F,\F\otimes\F^*> &=-||\bar\rd_A\F||^2 +||\rd_A\F||^2,\cr
||i\L F+\Ha\F\otimes\F^*-\Ha\t I||^2 &=||i\L F||^2+<i\L F,\F\otimes\F^*>-
\t<i\L F,I>\cr
&+{1\over 4} ||\F\otimes\F^*-\t I||^2,\cr
<i\L F,I> &={i\over {2\pi}}\int_X Tr(F,\o)\o^n=i\int_X TrF\w\o^{n-1},\cr}
}
and the K\"ahler identies,
\eqn\rrrv{
i[\L,\bar\rd_A]=\rd_A^*,\qquad -i[\L,\rd_A]=\bar\rd_A^*.
}

Finally we can obtain the invariant action
\eqn\lllv{\eqalign{
\CS=&||F||^2 +||D\F||^2 +{1\over 4}||\F\otimes\F^*
-\t I||^2-\t\int_X iTrF \w\o^{n-1}-\int_X Tr(F\w F)\w\o^{n-2}\cr
&+\int_M \c^{0,2}\w(\bar\ast\bar\rd_A\bar\p+\Fr{1}{16}[\f,\bar\ast\c^{0,2}])
+\x\w(\bar\ast\rd_A\p+\Fr{1}{8}[\f,\bar\ast\c^{2,0}]) \cr
& +\c\{ \L(\rd_A\bar\p +\bar\rd_A \p)+\Ha\bar\a\otimes\F^*+\Ha\F\otimes\a
+\Fr{1}{4}[\f,H]\} \o\w\bar\ast\o\cr
&+\eta\{\L(\bar\rd_A\p-\rd_A\bar\p)+\Ha\F\otimes\a-\Ha\F^*\otimes\bar\a\}
\o\bar\ast\o] \cr
& +\{\l((\bar\rd_A^*\rd_A +\rd_A^*\bar\rd_A)\f -\Ha\F\otimes[\F^*,\f]
+\Ha\F^*\otimes[\F,\f])\cr
& +\L(\bar\p\w[\l,\p]-\p\w[\l,\bar\p])\}\o\w\bar\ast\o]\cr}
}
under the following fermionic transformations;
\eqn\ttli{\eqalign{
s A' =\p,  \qquad  & s A'' =\bar\p,\cr
s\p =-\rd_A \f,\qquad & s\bar\p=-\bar\rd_A\f,\cr
s\F^* =\a, \qquad & s\F=\bar\a, \cr
s\a=-[\F^*,\f], \qquad & s\bar\a=-[\F,\f],\cr
s\f =0. \qquad &s\x=\bar\rd_A\F,\cr
s\c^{0,2}=4F^{0,2},\qquad &\c=i\L F+\Ha\F\otimes\F^*-\Fr{\t}{2}I_E,\cr
s\l =\eta,\qquad & s\eta=[\f,\l],\cr}
}

\newsec{Observables}

In this section, we hope to find the topological invariant observables.
The observables must be BRST invariant, not depend explicitly on the
metric, and not be written as $s$-exact, $s\r$.
These operators can be constructed as certain operators
\eqn\leei{
I(\CO)=\int_M\CW_{p,q}\w\CO,
}
where $\CW_{2-p,2-q}$, $(p,q=0,1,2)$, is a $(2-p,2-q)$-form on constructed
out of the fields and $\CO$ is $(p,q)$-form on $M$.
These are
\eqn\leeii{\eqalign{
& \CW^4_{0,0}=Tr\f^2,\cr
& \CW^3_{0,1}=Tr(2\bar\p\f),\cr
& \CW^2_{0,2}=Tr(\bar\p\w\bar\p+2F^{0,2}\f),\cr
& \CW^2_{1,1}=Tr(2\bar\p\w\p+2F^{1,1}\f),\cr
& \CW^3_{1,0}=Tr(2\p\f),\cr
& \CW^2_{2,0}=Tr(\p\w\p+2F^{2,0}\f),\cr
& \CW^1_{2,1}=Tr(2F^{2,0}\w\bar\p+2F^{1,1}\w\p),\cr
& \CW^1_{1,2}=Tr(2F^{0,2}\w\p+2F^{1,1}\w\bar\p),\cr
& \CW^0_{2,2}=Tr(2F^{2,0}\w F^{0,2}+F^{1,1}\w F^{1,1}).\cr}
}
These arise from the components $c^{(2,2)-i}$ of the second Chern class,
$c_2=Tr(\CF\CF)$, in the universal bundle.

There are the other observables obtained by Higgs fields,
\eqn\wwwi{\eqalign{
& \CW^0_{2,1}=\rd_A\F^*\w F^{1,1}+\bar\rd_A\F\w F^{2,0},\cr
& \CW^1_{2,0}=\rd_A\F^*\w\p +\a F^{2,0}+\bar\a F^{2,0},\cr
& \CW^1_{1,1}=\rd_A\F^*\w\bar\p +\bar\rd_A\F\w\p+\a F^{1,1}+\bar F^{1,1},\cr
& \CW^1_{0,2}=\bar\rd_A\F\w\bar\p+\a F^{0,2}+\bar\a F^{0,2},\cr
& \CW^0_{1,2}=\bar\rd_A\F\w F^{1,1}+\rd_A\F^*\w F^{0,2},\cr
& \CW^2_{1,0}=\rd_A\F^*\f +\a\p +\bar\a \p,\cr
& \CW^2_{0,1}=\bar\rd_A\F\f+\a\bar\p+\bar\a\bar\p,\cr
& \CW^3_{0,0}=\a\f+\bar\a\f.\cr}
}
These are obtained by $c'=Tr({\cal K}\CF)$ arised from the dimensional
reduction where ${\cal K}=\rd_A\F^* +\bar\rd_A\F +\a+\bar\a$\bau.

We can easely show that
\eqn\leeiv{
\rd \CW^4_{0,0}=s\CW^3_{1,0}, \qquad \bar\rd\CW^4_{0,0}=s\CW^3_{0,1},
}
and one finds recursively
\eqn\leev{\eqalign{
\rd\CW^{4-p-q}_{p,q}=s\CW^{4-p-1-q}_{p+1,q},& \qquad
\bar\rd\CW^{4-p-q}_{p,q}=s\CW^{4-p-q-1}_{p,q+1}\cr
 \rd\CW^0_{2,2}=0,  & \qquad \bar\rd\CW^0_{2,2}=0.\cr}
}

Then the integral $I(\CO)$ is BRST invariant, since
\eqn\rrr{
sI(\CO)=\int_M s\CW_{2-p,2-q}\w\CO=\int_{\g}\rd\CW_{2-(p-1),2-q}\w\CO=0
}
or
\eqn\rst{
sI(\CO')=\int_M s\CW_{2-p,2-q}\w\CO'=\int_M \bar\rd\CW_{2-p,2-(q-1)}\w\CO'=0.
}
Assume that $\CO=\rd\s$ or $\CO=\bar\rd\s$,
\eqn\jjj{\eqalign{
I(\CO) &=\int_M \CW_{2-p,2-q}\w\rd\s=(-1)^{p+q}\int_M
\rd\CW_{2-p,2-q}\w\s\cr
&=(-1)^{p+q}\int_M s\CW_{2-(p+1),2-q}\w\s=s((-1)^{p+q}\int_M
\CW_{2-(p+1),q}\w\s)\cr}
}
or
\eqn\jji{\eqalign{
I(\CO) &=\int_M \CW_{2-p,2-q}\w\bar\rd\s=(-1)^{p+q}\int_M
\bar\rd\CW_{2-p,2-q}\w\s\cr
&=(-1)^{p+q}\int_M s\CW_{2-p,2-(q+1)}\w\s=s((-1)^{p+q}\int_M
\CW_{2-p,2-(q+1)}\w\s).\cr}
}
The expectation values of such observables vanish
Therefore we can obtain the observables of topological BRST symmetry from
the de Rham cohomology classes of M.

Let us define the map
\eqn\nnn{
\P_{\CO}:\sum_{p+q=r} H^{p,q}(M)\to H^r(\CA^{1,1}\times \O^0),
}
which takes the form
\eqn\fff{
\P_{\CO}=\int_{M}\CW_{2-p,2-q}\w\CO
}
where $\CO$ is a $\bar\rd$-closed or $\rd$-closed $(p,q)$-form on $M$.
We can see that such a map corresponds to Donaldson map defineing over
cohomology.

\newsec{Conclusion}
We have shown that the vortex equations of $n$-dimensional K\"ahler
manifold are given by dimensional reduction of Einstein-Hermitian
equations of $(n+1)$-dimensional K\"ahler manifold.
The topological invariant action over the vortex sector leaded to
Yang-Mills-Higgs field theory with fermionic symmetry on closed
K\"ahler manifold. Two kinds of observables were obtained.
One are observables from the pure gauge fields and the others are
observables coupled the gauge fields with the Higgs fiels.
These facts follow also dimensional reduction.
However we need further to study the relation vortex moduli space over
$n$ dimension with  Einstein-Hermitian moduli space over $n+1$ dimension
in dimensional reduction.

\bigbreak\bigskip\bigskip\centerline{{\bf Acknowledgements}}
The author would like to express his gratitude to Professor J. H. Yee and
J. S. Park for helpful discussions. This work was
supported in part by the Center for Theoretical Physics (SNU), the Korea
Science and Engineering Foundation, the Ministry of Education and Daewoo
Foundation.

\listrefs
\bye